\documentclass{cimento}
\usepackage{graphicx}  
\title{A Critical Appraisal of Some Concepts Used in Neutrino Physics}
\author{Francesco Vissani, Manimala Mitra, Giulia Pagliaroli \from{ins:x}}
\instlist{\inst{ins:x}INFN, Gran Sasso, Theory Group, Assergi (AQ), Italy}
\begin{document}
\maketitle

\begin{abstract}
We examine the value of certain concepts highly regarded in the past decade, that
concern neutrino propagation, models for the leptonic mixing, interpretations of neutrinoless double beta decay and of SN1987A observations. We argue that it would useful to strengthen 
the role of the discussions among experts of neutrino physics,
regarding the hypotheses underlying the theoretical investigations.
\end{abstract}

\section{Superluminal neutrinos}
Superluminal neutrinos are no more needed \cite{15};
our goal, though, is to make a step back and 
examine how this concept arose. 
The ground was prepared by 
speculations on non-Einsteinian dispersion relations,
as the velocity is $\vec{v}=\partial E/\partial \vec{p}$.
\textrm{Gonzalez-Mestres '97}  proposed
$E^2=m^2+\left[\ \sin(p\ a)/a\ \right]^2$ for hadrons, 
with $a\sim {1}/{M_{\mbox{\tiny Planck}}}\equiv
\sqrt{G_N}$, 
arguing that the new kinematics can wipe out the GZK cutoff;
the 3-4$\sigma$ indication from AGASA is contradicted by AUGER.
\textrm{Amelino-Camelia, Ellis, Mavromatos, Nanopoulos, Sarkar '97} proposed 
$p^2=E^2 (1+\xi E/E_{\tiny\sf QG})$ for the photons, of which  ``quantum gravity'' was alleged. It implies $v=1-\xi E/E_{\tiny\sf QG}$ and thus a delay that depends on the energy; the 2.5$\sigma$ hint from MAGIC is excluded from HESS. \textrm{Coleman and Glashow '98} proposed $E_a= c_a \sqrt{p^2+(m_a c_a)^2}$ where $c_a\neq 1$  is a particle-depending constant; 
the interpretation along these lines of 
OPERA 2011 findings \cite{1} was 
criticized by many theorists.

\begin{figure}[b]
\begin{minipage}[b]{6.5cm}\label{cap}
\caption{\footnotesize Citations received by the study of neutrino velocity performed by 
MINOS 2007, that show that the upper bound was considered of limited interest till past year and that the subsequent analysis of OPERA triggered an outburst of interest. Note the relatively large number of papers published recently. From the NASA/ESO database, March 2012.\vskip1mm\tiny.}
\end{minipage}
\hfil
\includegraphics[width=0.50\textwidth,height=0.23\textwidth,angle=0]{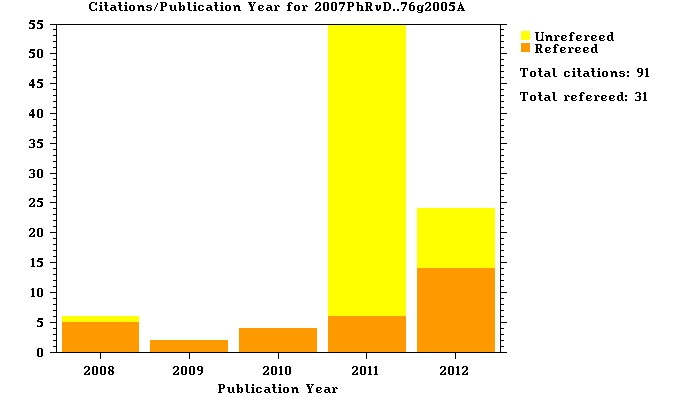}
\end{figure}

MINOS begun the recent campaign of measurement of neutrino velocity 
with these motivations \cite{menos}
{\footnotesize\em ``...theories have been proposed to allow some or all neutrinos to travel along ``shorcuts'' off the brane through large extra dimensions (5), and thus have apparent velocities different than the speed of light. Some of these theories (6-8) allow $|v-c|/c\sim 10^{-4}$ at neutrino energies of a few GeV.''}
Ref.~(6) is Ann.~Fond.~Broglie 31 (2006) 227 of \textrm{Volkov},  
Refs.~(7,8) are unpublished. 
Ref.~(5) 
by 
\textrm{Mohapatra} and 
\textrm{Smirnov}  
discusses `branes' and `extra dimensions' but does not mention `shortcuts'.
The paper {\em Sterile-active neutrino oscillations and shortcuts in the extra dimension}
by \textrm{P\"as, Pakvasa, Weiler,}
is not quoted in any of these works. 
The word `theories' used to introduce 
Refs.~(5-8) denotes respect, but 
does not mean that they have the  status, say,  
of QED, of relativity or of quantum theory. Note that we call `models' and not `theories'
the standard description of the Sun by \textrm{Bahcall} and the one of elementary particles  
by \textrm{Glashow, Weinberg, Salam}.

The concept of superluminal neutrinos became appealing in the past decade.
Various supporting arguments have contributed to the positive attitudes toward OPERA 2011 findings \cite{1}. E.g., the declarations 
of Petronzio on the Italian newspaper {\it Il Messaggero} (Sept 23, 2011) 
allude to  the `extra dimensions' often mentioned in the past {\it Piano Triennale} of INFN. 
Evidently theorists are not to be blamed for mistakes in experimental analyses,  moreover unpublished; the issue is however  that they have the responsibility of what is considered interesting and what it is being discussed.

\begin{figure}[t]\label{fig2}
\begin{minipage}[b]{5cm}
\caption{The area in lighter color is the distribution probability of the $\theta_{13}$ as found from the experimental analyses; the smaller curves and the arrows, instead, indicate various theoretical predictions. From \cite{2}, where references can be found.\vskip3mm\tiny.}
\end{minipage}
\hfil
\hskip3mm\includegraphics[width=0.53\textwidth,height=0.24\textwidth,angle=0]{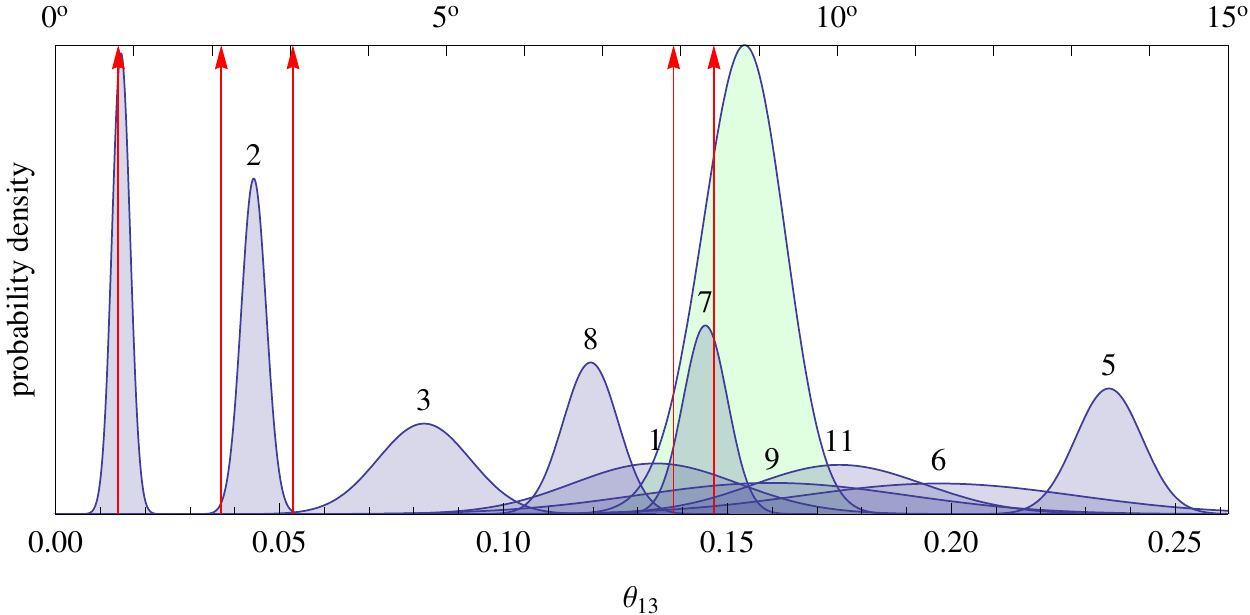}
\vskip-3mm
\end{figure}

\section{Leptonic mixing angles}

In the nineties, the solution of the solar neutrino anomaly preferred by 
many theorists was the small angle solution, now gone. 
Something similar happened with $\theta_{13}$.
E.g., \textrm{Harrison, Perkins, Scott} '02 posit a  mixing matrix  with $\theta_{13}=0$, termed ``tri-bimaxi\-mal'', that 
has had a significant impact on the general scientific discussion.\footnote{
It has been quoted more than 10 times by 
many prominent colleagues, including
\textrm{S King  {\bf\tiny\tiny 50}, 
Z-Z Xing  {\bf\tiny 39},  
E Ma  {\bf\tiny 35},
W Rodejohann  {\bf\tiny 29}, 
S Morisi  {\bf\tiny 28},   
G Altarelli  {\bf\tiny27} , 
L~Merlo  {\bf\tiny23}, 
S Antusch  {\bf\tiny22}, 
J Valle  {\bf\tiny21}, 
F Feruglio  {\bf\tiny20}, 
Y Koide  {\bf\tiny20},
C Hagedorn  {\bf\tiny18},
M Hirsch  {\bf\tiny18},
X-G He  {\bf\tiny18},
M Tanimoto  {\bf\tiny17},
R Mohapatra  {\bf\tiny15}, 
A Zee  {\bf\tiny14},
F Bazzocchi  {\bf\tiny13},
M-C Chen  {\bf\tiny13}, 
A Smirnov  {\bf\tiny13}, 
D~Meloni  {\bf\tiny13},
W Scott  
\& 
P Harrison  {\bf\tiny12},
W Grimus  {\bf\tiny12},
S Petcov  {\bf\tiny11} 
{\rm and}
P Frampton~{\bf\tiny11}}.
From inSPIRE database.}

The measurements do not corroborate similar positions and 
rule out many proposals, see Fig.~{\bf 2}.
The remaining proposals should be examined 
to assess their value.
E.g., Ref.~\cite{3} guessed the gross structure of the neutrino mass matrix $M_\nu$ by
one key parameter and describing the residual uncertainty  with a matrix of 
random numbers of $\mathcal{O}(1)$, namely
$
M_\nu \propto 
\mbox{diag}(\varepsilon,1,1)\cdot \mbox{\tt\footnotesize random} \cdot
\mbox{diag}(\varepsilon,1,1)
$. 
The best value -- much better than $\epsilon=1$ called ``anarchy'' -- was found to be 
$\varepsilon=\theta_C=13^\circ \sim \sqrt{m_\mu/m_\tau}=14^\circ$.
This value supported the large angle solution of the solar neutrino 
anomaly before it was confirmed, suggested a deviation of $\theta_{23}$ from the maximal value 
of similar size,  and yielded
$
\theta_{13}=12^\circ\!\pm 6^\circ$ 
in agreement with the recent $\theta_{13}$ measurements
(or $6^\circ\!\pm3^\circ$ with diagonal charged leptons).
 
Let us emphasize the peculiarities of this approach. Ref.~\cite{3} aims at an  
understanding of the mass matrix, rather than immediately postulating or discussing the 
mixing matrix. This seems a methodological merit,
 for the mixing matrix is derived by the mass matrix in gauge theories.
However, approaches as Ref.~\cite{3}  concern a  {\em class} of mass matrices: this calls for a  
more complete setup and, in fact, for a theory of the $\mathcal{O}(1)$  coefficients. 

The speculations starting from $\theta_{13}=0$, including most variants of tribimaximal mixings, have suggested that the conventional beams were not as appealing as the neutrino factories or beta beams; 
after \cite{db}
the value of this opinion is being reconsidered.

\begin{figure}[t]\label{fgh}
\begin{minipage}[b]{4.2cm}
\caption{Value of $m_{ee}$ from Klapdor results
\cite{kkk}; 
bound on $m_{\mbox{\tiny\sf lightest}}$ from cosmology \cite{8};
expectations from
{\tt dim.5} operators
and 3 flavor oscillations.
Left, normal hierarchy; 
right, inverted hierarchy.
Note the disagreement with the expectations.
\vskip5mm\tiny.}
\end{minipage}
\includegraphics[width=0.34 \textwidth,height=0.29\textwidth,angle=0]{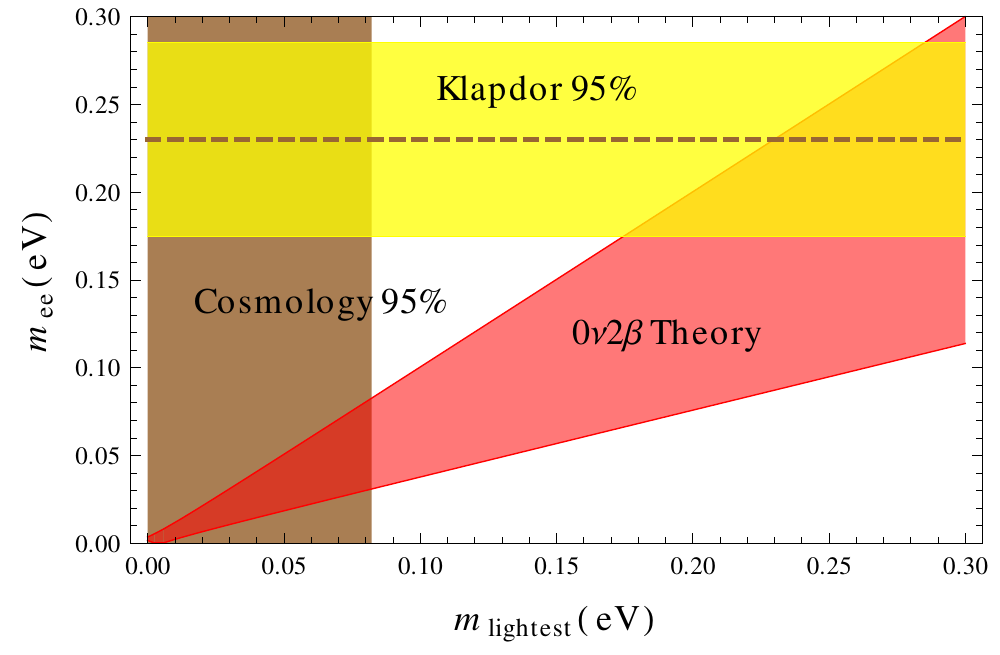}\hskip-1mm
\includegraphics[width=0.34\textwidth,height=0.29\textwidth,angle=0]{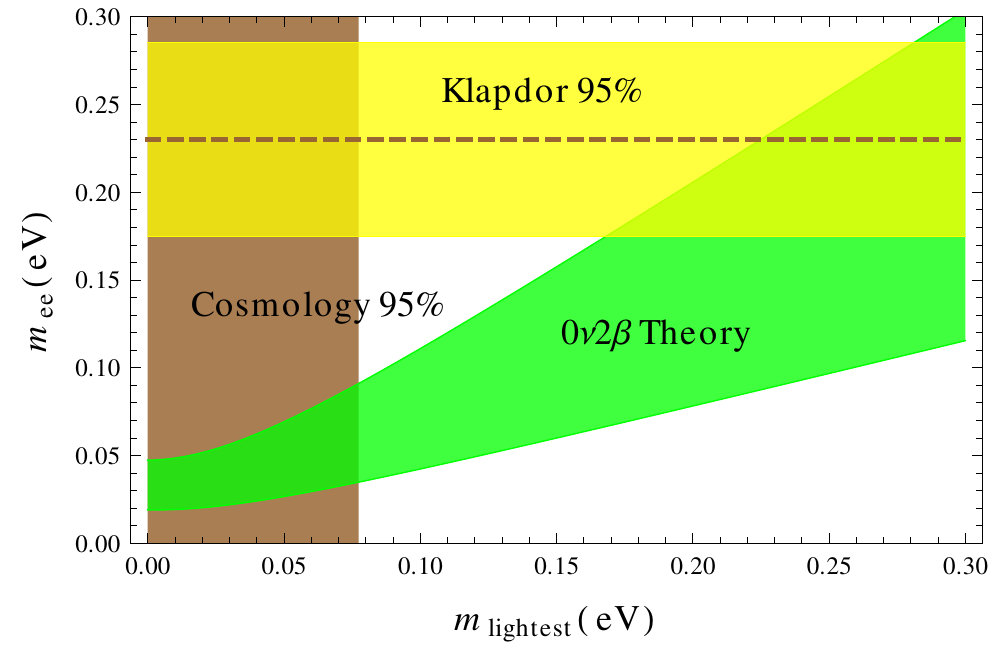}
\vskip-3mm
\end{figure}
\section{Neutrinoless double beta decay}

Various higher-dimensional operators, that respect the SU(3)$_c\times$SU(2)$_L\times$U(1)$_Y$ gauge symmetry, 
however violate the baryon and lepton numbers, as noted in \cite{4} and \cite{45};
$$
\delta\mathcal{L}= \frac{(\ell H)^2}{M}
+   \frac{\ell qqq}{M'^2}
+   \frac{(\ell q d^c)^2 }{M''^5}
\mbox{ with }
\left\{
\begin{array}{lr}
M<10^{11}\mbox{ TeV} & \mbox{for \tt dim.5}\\
M'>10^{12}\mbox{ TeV} & \mbox{for \tt dim.6}\\
M''>5\mbox{ TeV} & \mbox{for \tt dim.9}\\
\end{array}\right.
$$
The bounds on {\tt dim.5} comes from  
neutrino masses $m_\nu<0.1$ eV, the one on {\tt dim.6} from 
matter stability, and the one on {\tt dim.9} is from the test of lepton number violation that we 
discuss here.\footnote{Here 
we show just some representative operators. Note that if there are light sterile neutrinos, dark matter--or generally additional light states--more 
operators may be required, and that
a large effective mass could stem from small adimentional couplings $y$, e.g., $1/M=y^2/\mu$.
}
The transition $(A,Z)\to (A,Z+2)+2 e^-$, named neutrinoless double beta decay,  
can be induced by the operators of {\tt dim.5}, the one of {\tt dim.9}, and other ones. 
Which is the leading source of this process?
If the {\tt dim.5} operator, that provides us with Majorana neutrino mass terms, 
accounts for the observed three flavor oscillations and also 
dominates the transition,
the key quantity is the $e-e$ element of the neutrino mass matrix
$m_{ee}$, 
whose value depending on
the lightest neutrino mass 
can be calculated and usefully displayed  \cite{5} as shown in Fig.~{\bf 3}. 

The previous hypothesis is reasonable but does not apply  in general.
When the scale of  lepton number violation is low, the higher dimension operator 
can play a main role, and the connection with Majorana   
neutrino masses (i.e., with the {\tt dim.5} operators) is quite loose or just absent.
E.g.,  
neutrinoless double beta decay process
can be due to
sterile neutrinos below 10 GeV that 
explain neutrino masses~\cite{msv}.
Also in left-right extensions of the standard model 
that can be probed at the LHC, the 
{\tt dim.9} operators are relevant \cite{6}. 
Therefore, we cannot conclude  on logical grounds that we have 
a ``black-box theorem'', namely a necessary connection between 
the observation of neutrinoless double beta decay and a Majorana 
nature of the ordinary neutrinos. Rather, we can say that under reasonable conditions 
(if the higher order operators play 
no role, in absence of other light neutrinos, etc.) we have 
quantitative correlations 
as shown in Fig.~{\bf 3}.

Similar reluctance to analyze critically the views to which we are accustomed can be perceived also from the language. When we speak of neutrino{\em less} double {\em beta} decay, we use a terminology for initiates and define a reaction for the {\em absence} of neutrinos, which is quite repulsive to common 
sense -- even if it draws an analogy with the double beta decay and it 
recalls the absence of 
``missing energy''. 
Another useful description of the same is \textbf{creation of electrons in a nuclear transition}, that emphasizes the violation of the lepton number, rather than alluding to a theoretical interpretation in terms of virtual \textrm{Majorana} neutrinos--or in modern terms, the dominance of {\tt dim.5} operators. Moreover, such an alternative description can be explained also to laymen, 
it shows that the process is as important as proton decay
and suggests connections with leptogenesis.

\begin{figure}[t]
\begin{minipage}[b]{7.2cm}
\caption{\footnotesize 
Temporal distribution of the events of 
Kamiokande-II (KII), Baksan (BAK) and IMB. Compatibly with the errors 
on the absolute time of KII and BAK, 
the beginning has been set to be the same.
The 
rapid accumulation  of events in the first second is 
evident from the data: 6 events in Kamiokande-II, 3 in IMB, 2 in Baksan. The vertical line is where 
there are half of the events. 
From~\cite{11}.\vskip10mm\tiny.}
\end{minipage}
\vspace{2mm}
\includegraphics[height=0.33\textwidth,width=0.43\textwidth,angle=0]{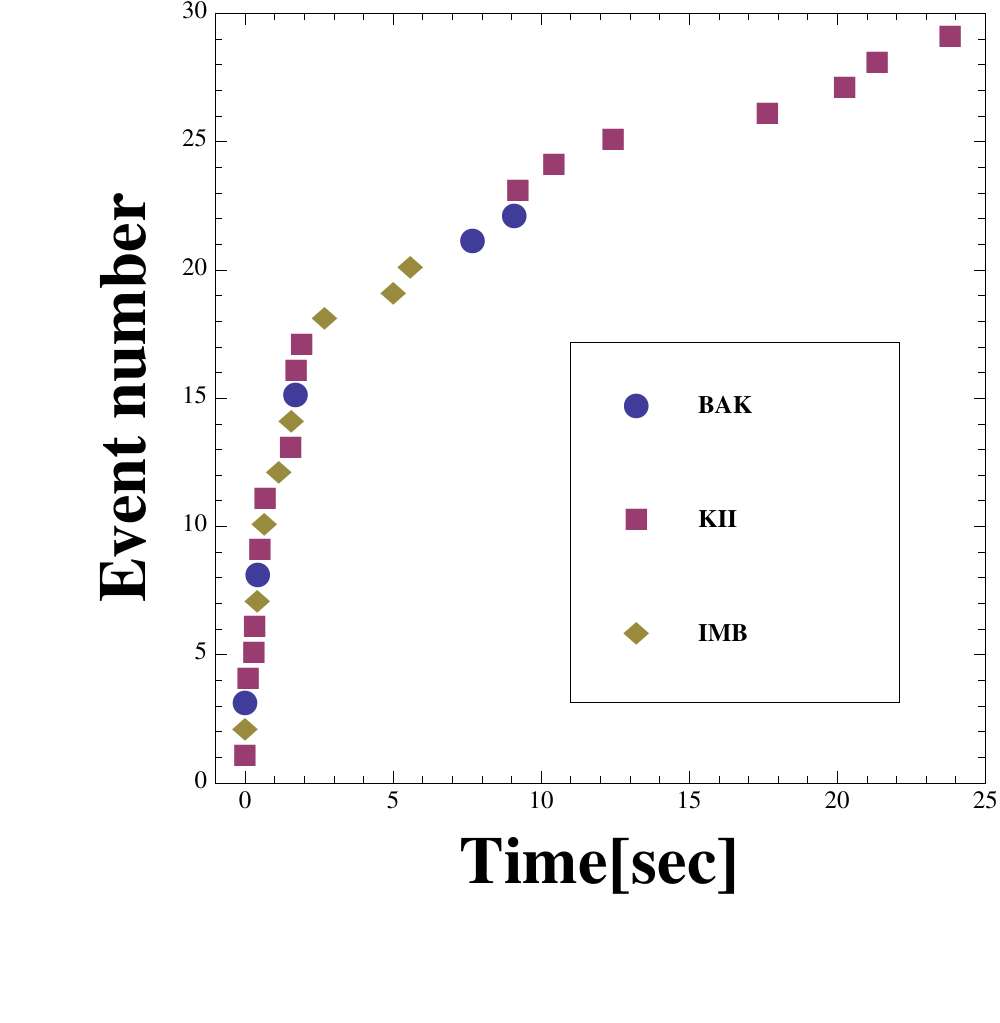}
\vskip-12mm
\end{figure}

\section{\bf Interpretations of SN1987A observations}

The observations of SN1987A by Kamiokande-II, IMB and Baksan have begun a new chapter of astronomy. The main discussion of the astrophysical aspects lasted few years; the subsequent discussion of supernova neutrinos has proceeded, quite irrespectively of the interpretation of the SN1987A observations. The discussions in particle physics instead lasted much longer, and concerned mostly  
neutrino properties (initially, neutrino masses; later, neutrino mixings; more 
recently, exotic aspects). Based on \cite{11}, we would like to emphasize 
some attitudes of the discussion, that illustrate its limitations:

$(i)$~A diffuse opinion has been--and it is--that SN1987A was `non-standard'. E.g., it has been repeated that the average energy of the events observed from SN1987A is {\em too}  low. However, the most recent simulations support lower energies -- which suggests that the uncertainties are not understood.
$(ii)$~\textrm{Smirnov, Spergel \& Bahcall} have discussed in Phys.Rev.~D49 (1994) 1389 whether SN1987A excludes large lepton mixing. A posteriori, the answer is evident, but the point to reiterate is that questions like these cannot be addressed before knowing the astrophysical uncertainties. 
$(iii)$~The previous two issues regard the energy distribution of the events; curiously, the 
meaning of the temporal distribution of the events, shown in 
Fig.~{\bf 4}, has been discussed thoroughly only quite recently.
$(iv)$~The observations of Baksan have been often ignored. 
Similarly, it is not clear whether the discussion of LSD findings was as complete as possible.
$(v)$~Important pending questions, such as the existence of a compact remnant (neutron star?), or of multiple neutrino emissions, have received--and receive--only a marginal attention. 

Perhaps, now that we know a  lot on neutrino properties, the scope of the discussion of SN1987A events will widen.

\section{Discussion}

We considered from various points of view certain theoretical
concepts pertinent to 
superluminal neutrinos; models for $\theta_{13}$; 
neutrinoless double beta decay; SN1987A.
In several cases, ideas that became popular and attracted consensus  (as measured by conferences, publications, citations) do not seem to correspond to valid concepts. A natural question is whether we can  
avoid this type of polarization.

Let us examine the issue in general terms. Physics
requires an extensive use of deductive (or analytical) methods -- here is where mathematics acts as a very effective tool -- but it needs also to apply inductive procedures. The new concepts, or the attitudes of the scientific discussions, belong mostly to inductive aspects of the method, and they should 
be subjected to critical attention in order to function properly. 
This correspond to the {\em pars destruens} of \textrm{Bacon}'s inductive method
and 
can be summarized with  \textrm{Newton}'s words {\em hypotheses non fingo}.

The shortage of fresh data is not the only problem that should worry us. We believe that concepts, hypotheses and results should undergo critical examinations, and in our humble opinion we are called to make more efforts in this sense.  In the same spirit, we think that open and frank scientific discussions among experts ought to play a more important role in neutrino physics.  Activities like these are worthwhile even if (or just because) they may lead to opinions in partial contrast with current trends/hot topics.

\acknowledgments
F Vissani thanks the Organizers of \textrm{IFAE 2012} for the invitation and 
{\rm F~Aharonian,
F~Ferroni,
M~Goodman,
G~Senjanovic,
P~Strolin,}
and
{\rm L~Toscano}
for useful discussions.

\section*{Two months after -- note added}
The healthy state of experimental neutrino physics is unquestionable:
RENO collaboration has released data that corroborate Daya 
Bay results on $\theta_{13}$;
also, a null result from EXO-200
excludes the largest values of $m_{ee}$ compatible 
with Klapdor's findings.

  \end{document}